\documentclass[a4paper, fontsize=11pt]{article}

\renewenvironment{abstract}
               {\list{}{\rightmargin\leftmargin}%
                \item[\hspace*{1cm}\small\textbf{Abstract ---}]\relax}
               {\endlist}

\usepackage[top=.8in, bottom=.8in, left=.8in, right=.8in]{geometry}
\usepackage[fleqn]{amsmath}
\usepackage{amssymb}
\usepackage{enumerate}
\usepackage{amsthm}
\usepackage[pdftex]{graphicx}
\usepackage{balance}
\usepackage{fancyhdr}
\usepackage{sidecap}
\usepackage{appendix}
\usepackage{float}
\usepackage{scrextend}
\usepackage{changepage}
\usepackage{subfigure}
\usepackage{endnotes}

\let\footnote=\endnote

\newtheorem*{Postulate*}{}

\begin{document}

\title{Reply to Hofer-Szab\'{o}: the PBR theorem hasn't been saved}

\author{Marcoen J.T.F. Cabbolet\\
        \small{\textit{Center for Logic and Philosophy of Science, Vrije Universiteit Brussel\footnote{email: Marcoen.Cabbolet@vub.be}}}}
\date{}

\maketitle
\vfill
\begin{abstract} \footnotesize Recently, in \emph{Found. Phys.} \textbf{53}: 64 (2023), it has been argued that there is no reality to the PBR theorem. In \emph{Found. Phys.} \textbf{54}: 36 (2024), Hofer-Szab\'{o} has commented that the argument is flawed and that PBR theorem remains in tact. Here we reply to Hofer-Szab\'{o} by showing that his counterargument does not hold up, concluding that the PBR theorem has been disproved.
\end{abstract}

\noindent The PBR theorem alleges that $\psi$-epistemic quantum mechanics (QM) gives rise to predictions that are inconsistent of those of $\psi$-ontic QM if we do a special joint measurement on a bipartite system consisting of two subsystems, each prepared in one of two states that satisfy some assumptions \cite{Pusey}.  More precisely, if we prepare the bipartite system in one of four quantum states $|\ \psi_1\ \rangle  = |\ 0\ \rangle|\ 0\ \rangle$,  $|\ \psi_2\ \rangle  = |\ 0\ \rangle|\ +\ \rangle$, $|\ \psi_3\ \rangle  = |\ +\ \rangle|\ 0\ \rangle$ , or $|\ \psi_4\ \rangle  = |\ +\ \rangle|\ +\ \rangle$, and if we assume that we can do a measurement that projects on the four PBR states  $|\ \xi_1\ \rangle  \bot |\ \psi_1\ \rangle $, $|\ \xi_2\ \rangle  \bot |\ \psi_2\ \rangle $, $|\ \xi_3\ \rangle  \bot |\ \psi_3\ \rangle$, or $|\ \xi_4\ \rangle  \bot |\ \psi_4\ \rangle $ , then the PBR theorem alleges that $\psi$-epistemic QM predicts that there is a nonzero probability that the bipartite system prepared in the quantum state $|\ \psi_n\ \rangle$ upon measurement can end up in the PBR state  $|\ \xi_n\ \rangle  \bot |\ \psi_n\ \rangle $ if there is an ontic state of a subsystem that corresponds to both pure quantum states $|\ 0\ \rangle$ and $|\ +\ \rangle$. This projection $|\ \psi_n\ \rangle  \rightarrow |\ \xi_n\ \rangle $ is impossible in the framework of $\psi$-ontic QM, in the framework of which there is no ontic state of a subsystem that corresponds to both pure quantum states $|\ 0\ \rangle$ and $|\ +\ \rangle$.

In \cite{Cabbolet}, however, it has been argued that there is no reality to the PBR theorem by showing that for an ensemble of bipartite systems made up of two bolts, which satisfies all assumptions of the PBR theorem, it is not only the case that we do not obtain any of the predictions of $\psi$-epistemic QM that are inconsistent with $\psi$-ontic QM: in addition, the entangled measurement, for which an experimental design has been suggested, has no outcome in 25\% of the cases. %This argument is expanded upon in \cite{Addendum}, where the PBR theorem is disproven by a generic counterexample: it is shown that there is a generically defined model of an ensemble of bipartite systems, such that all assumptions of the PBR theorem are valid in the model, but such that there is zero probability for a projection $|\ \psi_n\ \rangle  \rightarrow |\ \xi_n\ \rangle $ to occur, and such that there are cases in which projection to one of the four PBR states does not occur at all.

In \cite{Szabo}, Hofer-Szab\'{o} argues that the argument against the PBR theorem in \cite{Cabbolet} is flawed, because the suggested experiment is not the entangled measurement of the PBR theorem  :
\begin{quote}
  ``{\it Cabbolet constructs a measurement which in the 1/4 of the cases does not yield an outcome. However, it turns out that this measurement
is not entangled and hence is not the entangled measurement of PBR. ... Thus, Cabbolet’s claim on the nonreality of the PBR theorem is unjustified}''
\end{quote}
Now it would certainly be an interesting discussion to establish whether the suggested measurement is or is not the entangled measurement of the PBR theorem. But even if it turns out that Hofer-Szab\'{o} is right, this is \emph{irrelevant}: the crux is that on the one hand, we can model all assumptions of the PBR theorem in the framework of the ensemble of pairs of bolts, while on the other hand there is, for any of the four quantum states  $|\ \psi_n\ \rangle$, zero probability that a member of the ensemble prepared in the quantum state $|\ \psi_n\ \rangle$ upon measurement ends up in the PBR state $|\ \xi_n\ \rangle  \bot |\ \psi_n\ \rangle $ \emph{regardless of how we design the measurement}. This argument is expanded upon in \cite{Addendum}, where the PBR theorem is disproven by a generic counterexample: it is shown that there is a generically defined model of an ensemble of bipartite systems, such that all assumptions of the PBR theorem are valid in the model, but such that there is zero probability for a projection $|\ \psi_n\ \rangle  \rightarrow |\ \xi_n\ \rangle$ to occur, and such that there are cases in which projection onto one of the four PBR states does not occur at all.
\newpage

Importantly, to disprove the PBR theorem we don't need to construct a model in which the measurement always projects onto one of the four PBR states. Namely, the derivation in \cite{Pusey} of the nonzero probability of the event that the bipartite system upon measurement ends up in a PBR state orthogonal to the quantum state in which it had been prepared \textbf{does not} lean on the assumption that the measurement always projects onto one of the four PBR states: this derivation merely uses the assumption that there is at least one ontic state of a subsystem that corresponds to both pure quantum states $|\ 0\ \rangle$ and $|\ +\ \rangle$ , so that the measurement device can ``see'' the bipartite system in any of the four states  $|\ \psi_n\ \rangle $ when both subsystems are in such an ontic state, and the assumption that one of the possible states upon measurement is the PBR state orthogonal to the quantum state in which the system has been prepared. In other words, the argument underlying the PBR theorem also applies if we prepare the bipartite system in a quantum state $|\ \alpha\ \rangle|\ \beta\ \rangle \in \{|\ 0\ \rangle|\ 0\ \rangle, |\ 0\ \rangle|\ +\ \rangle, |\ +\ \rangle|\ 0\ \rangle, |\ +\ \rangle|\ +\ \rangle\}$, and we then do a measurement that projects on the states $|\ \alpha\ \rangle|\ \beta\ \rangle$, $|\ \alpha^\bot\ \rangle|\ \beta^\bot\ \rangle$ (with $|\ \alpha^\bot\ \rangle$ and $|\ \beta^\bot\ \rangle$ being the orthogonal counterparts of $|\ \alpha\ \rangle$ and $|\ \beta\ \rangle$, respectively), and $\left( |\ \alpha\ \rangle|\ \beta^\bot\ \rangle + |\ \alpha^\bot\ \rangle|\ \beta\ \rangle \right)/\surd2$ (being the PBR state orthogonal to $|\ \alpha\ \rangle|\ \beta\ \rangle $): we then get the prediction that there is a nonzero probability that we find the bipartite system upon measurement in the PBR state  $\left( |\ \alpha\ \rangle|\ \beta^\bot\ \rangle + |\ \alpha^\bot\ \rangle|\ \beta\ \rangle \right)/\surd2$.

To spell it out for the case $|\ \alpha\ \rangle|\ \beta\ \rangle = |\ 0\ \rangle|\ 0\ \rangle$, consider that the bipartite system has been prepared in the quantum state $|\ 0\ \rangle|\ 0\ \rangle$. Then there is a nonzero probability that both subsystems are in an ontic state corresponding to both pure quantum states $|\ 0\ \rangle$ and $|\ +\ \rangle$. If we then do a measurement that projects on the quantum states $|\ 0\ \rangle|\ 0\ \rangle$, $|\ 1\ \rangle|\ 1\ \rangle$, and $\left( |\ 0\ \rangle|\ 1\ \rangle + |\ 1\ \rangle|\ 0\ \rangle \right)/\surd2$, there is thus a nonzero probability that the measurement device ``sees''  the bipartite system in one of the quantum states $|\ 0\ \rangle|\ +\ \rangle$, $|\ +\ \rangle|\ 0\ \rangle$, or $|\ +\ \rangle|\ +\ \rangle$, for each of which the inner product with the PBR state $\left( |\ 0\ \rangle|\ 1\ \rangle + |\ 1\ \rangle|\ 0\ \rangle \right)/\surd2$ is nonzero: there is, then, thus a nonzero probability that we find the bipartite system upon measurement in that PBR state $\left( |\ 0\ \rangle|\ 1\ \rangle + |\ 1\ \rangle|\ 0\ \rangle \right)/\surd2$. That shows that the PBR theorem does not lean on the assumption that there has to be a measurement that projects onto one of the four PBR states.

Even stronger, suppose we prepare just one subsystem of the bipartite system in the quantum state $|\ 0\ \rangle$, and we then do a measurement that projects on the quantum states  $|\ 0\ \rangle$ and $|\ 1\ \rangle$. Then there is a nonzero probability that the simple system is in an ontic state corresponding to both pure quantum states $|\ 0\ \rangle$ and $|\ +\ \rangle$. So, there is a nonzero probability that the measurement device ``sees'' the simple system in the quantum state $|\ +\ \rangle$. Given the inner product $\langle\ 1\ |\ +\ \rangle = 1/\surd2$, there is then a nonzero probability that the simple system upon measurement will be found in the quantum state $|\ 1\ \rangle$, which is orthogonal to the quantum state $|\ 0\ \rangle$. That shows that the argument underlying the PBR theorem even applies to simple systems.\\
\ \\
Concluding, it is herewith considered proven that Hofer-Szab\'{o}'s counterargument does not hold up: even if the measurement suggested in \cite{Cabbolet} is not the entangled measurement meant in \cite{Pusey}, this doesn't save the PBR theorem---it has to be retracted. The crux is, as shown in \cite{Addendum}, that the derivation in \cite{Pusey} of the nonzero probability for a projection $|\ \psi_n\ \rangle  \rightarrow |\ \xi_n\ \rangle $  leans on a tacit assumption that is false.

\end{document}